\documentclass[a4paper]{llncs}
\usepackage{amssymb}
\usepackage{amsmath}
\usepackage{array}
\usepackage{latexsym}
\usepackage{graphicx}
\usepackage{epsfig}
\usepackage{amsfonts}
\usepackage[mathscr]{eucal}

\newcommand{\eps}{\varepsilon}
\newcommand{\R}{\mathbb{R}}
\newcommand{\N}{\mathbb{N}}

\newcommand{\wei}{\mathrm{weight}}
\newcommand{\ba}[1]{\begin{array}{#1}}
\newcommand{\ea}{\end{array}}

\newcommand{\n}{\underline{n}}

\title{Communication regimes in opinion dynamics: Changing the number of communicating agents}
\author{Diemo Urbig\inst{1} \and Jan Lorenz\inst{2}
\thanks{The second author received financial funding by Friedrich-Ebert-Stiftung, Bonn, Germany.} 
}                     
\institute{Humboldt University of Berlin, Department of Computer Science, Berlin, Germany \and University Bremen, Department of Mathematics, Bremen, Germany}

\date{}

\begin{document}

\thispagestyle{empty}
\maketitle

\vspace{-7cm}
\begin{small}
\noindent Diemo Urbig and Jan Lorenz (2004): Communication regimes in opinion dynamics: Changing the number of communicating agents. {\em Proceedings of the Second Conference of the European Social Simulation Association (ESSA)}, September 16-19, Valladolid, Spain, 2004 (ISBN 84-688-7964-9). 
\end{small}

\vspace{+5cm}
\begin{abstract}
This article contributes in four ways to the research on
time-discrete continuous opinion dynamics with compromising
agents. First, communication regimes are introduced as an
elementary concept of opinion dynamic models. Second, we
develop a model that covers two major models of
continuous opinion dynamics, i.e. the basic model of Deffuant and
Weisbuch as well as the model of Krause and Hegselmann. To combine
these models, which handle different numbers of communicating
agents, we convert the convergence parameter of Deffuant and
Weisbuch into a parameter called self-support. Third, we present
simulation results that shed light on how the number of
communicating agents but also how the self-support affect
opinion dynamics. The fourth contribution is a theoretically
driven criterion when to stop a simulation and how to extrapolate to
infinite many steps.
\end{abstract}

\section{Introduction}
Opinion dynamic (OD) models describe the process of opinion formation 
in groups of individuals. We focus on continuous opinion dynamics 
(COD) with compromising agents in a time-discrete world. This 
implies that an opinion is a continuous value between zero and one. 
In every time step, each agent adapts his opinion towards the 
opinions of a set of perceived agents,
while the new opinion is between minimum and maximum of the own
and all perceived opinions (compromising). A common feature among 
many models of continuous opinion dynamics is bounded confidence, 
which describes the fact that opinions far away from
an agent's own opinion do not exhibit any influence on this agent.

Two models of continuous opinion dynamics have received much
attention. One the one hand, the model of Deffuant and Weisbuch
(DW model) 
\cite{DeffuantEtal2002,DeffuantEtal2000,WeisbuchEtal2001,WeisbuchEtal2002}
and on the other hand the model of Krause and Hegselmann (KH
model) \cite{HegselmannKrause2002}. A
fundamental difference between these two models is the number of
agents that communicate. In each time step in the DW model two randomly chosen
agents mutually perceive their opinions, while in the KH model all agents
perceive all other agents. The same tendency towards extreme
models regarding the number of communicating agents can be found
in the related literature on discrete opinion dynamics (see for
instance \cite{SchweitzerHolyst2000} but also
\cite{Fortunato2004}).

The difference between the DW model and the KH model is in a
dimension that Hegselmann calls the update
mechanism \cite{Hegselmann2004}. We would label it the communication regime, which
relates the dimension more to social reality than to models.
Social reality may restrict the process of communication between
agents such that communication between all agents at the same time
(as done in the KH model) but also communication between
two randomly chosen agents (as done by the DW model) appear as
extreme cases. Apart of restricting the number of
communicating agents we might think of social or physical networks
bounding the set of agents who could potentially affect an
agent. There are some simulation results about opinion dynamics on
social networks, see for instance
\cite{AmblardDeffuant2004,Fortunato2004,StaufferEtal2004}.
But static social networks are not the only restriction we may
think of. It could make a difference if we first communicate with
our mother and then with our friends or vice versa. Hence, also
temporal aspects can play a major role. In other words, we want to
introduce rules about who is communicating with whom at what time.
We call such rules and restrictions the {\em communication regime}
of an opinion dynamics model. We call it 'regime' because it is
independent of the opinions the agents have; it is a parameter of
the model. The communication regime includes the underlying social
network but also how this may change over time. In a mathematical
sense the communication regime is a (temporal) sequence of networks of who
perceives whose opinion. It can be treated as module of OD
models.

Another module of opinion dynamic models is what we call the {\em
mental model}. The mental model introduces rules how agents adapt
their opinions based on a set of other agents' opinions. It models
perceptual or information processing biases, e.g. bounded confidence. 
By many parallel and sequential
communications these biases are multiplied in a group such that
new sometimes even more complex group dynamics can be observed. 
Increasingly the mental model is treated as a module of OD models
independently of the communication regimes. For instance the mental model of the
basic DW model is put on network structures (see
\cite{AmblardDeffuant2004,StaufferEtal2004}), or the KH model is developed
with a mental model that captures discrete opinions instead of
continuous opinions (see \cite{Fortunato2004}).

In the section \ref{sec:genmod} we provide a general model of
continuous opinion dynamics that extends the general model presented in 
\cite{HegselmannKrause2002} by explicitly including communication
regimes. In section \ref{sec:examples} we provide examples how
communication regimes can be used to capture parts of other
opinion dynamic models and how they can be used to model different
kinds of social networks. The major difference between the basic
model of Deffuant and Weisbuch on the one hand and the model of
Krause and Hegselmann on the other hand is the communication
regime. Hence, our general model should be able to provide a model
that contains both models. Such a model is provided in the section
\ref{sec:smallestmodel}. As a prerequisite we unify the mental
models; thereby we re-define the convergence parameter $\mu$ (now
called self-support) of the DW model and introduce this idea into
the KH model. Section \ref{sec:simulation} summarizes simulations
that show how the differences in the communication regime 
influences the dynamics. We also show that the impact
of self-support is moderated by the communication regime.

\section{A Generalized Model of Continuous Opinion Dynamics with Communication
Regimes}\label{sec:genmod}

In this section we develop a generalized COD model. Its mental model is based on
weighted averaging. Communication regimes are implemented by
sequences of matrices. We formulate the model as a time-discrete
dynamical system written in 'matrix language', where the agents are
associated with the indices of the matrix. As we will see later this formalization of communication regimes also supports the analysis of social networks in opinion dynamics.
Our general model is a modification of the general model that is presented in \cite{HegselmannKrause2002}.

\paragraph{Agents and their opinions}
Consider a set $\n:=\{1,\dots,n\}$ of agents. We call the vector
$X(t) \in \R^{n}$ an {\em opinion profile}. $X_i(t)$ for $t \in
\mathbb{N}_0$ represents the opinion of agent $i$ at time $t$. The
initial opinion profile is given by $X(0)$.

\begin{definition}[communication regime]
A sequence $C(t) \in \{0,1\}^{n\times n}$ for $t\in \N$ is called
a {\em communication regime} if the diagonal is positive in every
matrix.
\end{definition}

A communication regime is thus a sequence of $(0,1)$-matrices, which might also be called communication matrices. If
$c_{ij}(t)=1$ then agent $i$ perceives agent $j$ at time step $t$.
Hence, $C(t)$ is the adjacency matrix of the network of who
perceives whom at time step $t$, i.e. the $i$-th row in $C(t)$ marks the
{\em set of perceived agents} of agent $i$. The
positive diagonal means that self-communication should always be
possible. For the set of perceived agents we define the mental model as an
abstract weighted averaging rule. For a discussion and analysis of
different abstract averaging rules in opinion dynamics see 
\cite{HegselmannKrause2003}.

\begin{definition}[averaging rule] \label{def:avru}
Let $C(t)$ be a communication regime and $i\in\n$ be an agent. A
function $\wei_i: \n \times \R^n  \times \mathbb{N} \to  [0,1]$ is called an 
\emph{averaging rule} if it holds
\begin{eqnarray*}
\forall t \in \mathbb{N} \;\;\; \forall X\in\R^n \;\;\; &:& \;\;\;
\sum_{j=1}^n\wei_i(j,X(t),t) = 1 \\
\forall t \in \mathbb{N} \;\;\; \forall X\in\R^n \;\;\; \forall j \in \n &:& \;\;\; c_{ij}(t) = 0
\Longrightarrow \wei_i(j,X(t),t) = 0
\end{eqnarray*}
\end{definition}

An averaging rule $\wei_i(j,X(t),t)$ shows how much weight agent $i$
puts on opinion of agent $j$ at time step $t$. The first constraint implements a
normalization of the weights an agent puts on all agents
(including himself), while the last constraint ensures that agent
$i$ can only put weights on actually perceived agents. The weight function can 
implement many different approaches. Based on the weight function we can
complete the construction of the mental model. Thereby we adopt
the idea of the confidence matrix as it is also done by
\cite{HegselmannKrause2002,Lorenz2003c}; but additionally we require
this matrix to be along with the communication regime. Finally we
can recursively define the process of opinion formation.

\begin{definition}[confidence matrix]
Let $C(t)$ be a communication regime, $X(t)$ an opinion profile
and let $\wei_i$ be an averaging rule for all $i\in\n$, then
\[
A(t,X(t),C(t))_{[i,j]} := \wei_i(j,X(t),t)
\]
is called a {\em confidence matrix\/}.
\end{definition}

Due to definition \ref{def:avru} for every time step $t$ matrix $A(t,X(t),C(t))$,
the communication matrix, is
row-stochastic. Further on, it holds that $A(t,X(t),C(t)) \leq C(t)$. This
represents, that the mental model is restricted by the
communication regime. Finally we define the process.

\begin{definition}[COD process]
Let $C(t)$ be a communication regime, $X(0) \in \R^{n}$
be an initial opinion profile and $\wei_1, \dots, \wei_n$
averaging rules for all agents in $\n$. The {\em process of  
continuous opinion dynamics} is a sequence of opinion profiles
$(X(t))_{t\geq 0}$ recursively defined through
\[ X(t+1) = A(t,X(t),C(t))X(t).\]
\end{definition}

\section{Examples for Communication Regimes} \label{sec:examples}

Our general model explicitly introduces the concept of
communication regimes. Some examples show the variety of structures
that may be modeled.

\paragraph{Communication in groups}
If $C$ is the unit matrix then there is no communication at all.
If contrary $C$ contains only ones then it is the communication
regime as modeled in \cite{HegselmannKrause2002}, i.e.
every agent perceives every other agent. We can see this
communication regime as a commission trying to pool the commission
members' opinions where every agent gets the opinion of every
other agent before starting to rethink about his own new opinion.
Between 'no communication' and 'communication with all agents' we
can define communication in groups of $m$ agents, where all agents
in a group perceive only the opinions of their group members. We
may think of people going to lunch each day and discuss with their
neighbors. The communication regime of the DW model is of this
type with a group size of two. We may think of it as a society
where in every step one agent spontaneously decides to phone
another agent out of his phone book to try to compromise with. We
summarize all these communication regimes as {\em
$m$-communication regimes}, where parameter $m$ characterizes the
group size. For binary opinions such a parameter was investigated
in a model by \cite{Galam2002}.

\begin{definition}[$m$-communication regime]
Let $\n$ be a set of agents and $m\leq n$. We call a communication
regime $C(t)$ \emph{$m$-communication regime} if for all $t\in\N$
it holds that $C(t)$ is symmetric and there are exactly $m$ rows
containing exactly $m$ ones and the rest zeros. All other rows
contain only zeros, except the diagonal entries, which are ones.
\end{definition}

For a system with $n$ agents there is only one $n$-communication
regime, but there may be many different $2$-communication regimes.
Models with $m$-communication regimes with $m$ less than $n$
frequently include a random choice of the actually applied
communication regime. For each matrix from the communication
regime $m$ agents are randomly selected to communicate. We call
these classes of communication regimes \emph{random
$m$-communication regimes}. The random $n$-communication regimes
is a special case not including any random choice. In the DW model
the communication regime is a random $2$-communication
regime; thus for each step of the opinion dynamics there is a
random pairwise communication. Compared to the KH model this adds
an additional source of randomness and perhaps additional variance
to the model. The DW model and the KH model are extreme cases
of this class of communication regimes, i.e. for $m=2$ and $m=n$,
respectively. By formalizing random $m$-communication regimes we
implement an intuition that is also mentioned in
\cite{Hegselmann2004}.

\paragraph{Social networks}
If we only apply $C(t) = C$ then we get the concept of networks as
applied by French \cite{French1956}, with $C$ being the adjacency
matrix of the social network. Matrix $C$ formalizes the structure
of a static social network. If we are less restrictive and only
require $C(t) \leq C$ for all $t$ then we define the {\em
underlying social network} as the smallest $C$ that satisfies
$C(t) \leq C$. If communication partners are randomly chosen
according to a social network represented by a matrix $C$ then the
underlying social network is $C$. Such random communication on a
network $C$ is implemented for instance by \cite{StaufferEtal2004}.

\paragraph{Spatial or temporal structures}
Another very specific example is $c_{ij}=1$ for all $i,j \in \n$
with $|i-j| \leq 1\}$; this is "communication in a line", where line 
refers to a spatial concept. Many more complex static communication 
regimes as for instance the Moore-neighborhood or other spatial arrangements fit
into this concept of communication regimes.

Since $C(t)$ depends on $t$ the structure who perceives whom may change over time and definitions of time schedules and frequencies are possible, e.g. close neighbors were perceived more frequently than agents of greater distance. This also allows for the definition of multi-layer social networks resulting in different communication frequencies as it can be found for instance in \cite{ChattoeGilbert1999}.

\section{The Smallest Model covering DW and KH} \label{sec:smallestmodel}

In this section we present a general model that explicitly
addresses the modeling of communication regimes. In the previous
section we saw that the DW model as well as the KH
model apply random $m$-communication regimes, while they differ
only in the parameter $m$. In this section we want to utilize this
for a unification of these two influential models. To find a
parameterized model covering both, DW model (with globally uniform
uncertainty and without relative agreement) and KH model, we have to make
the communication regimes and the mental models compatible; this
means we have to define an appropriate communication regime and an
averaging rule.

Since both models apply random $m$-communication regimes, a
general model containing both specific model therefore only needs
to incorporate such communication regimes with an integer
parameter $m$ less or equal to the number of agents $n$. Thus we
restrict us in this particular model to random $m$-communication
regimes, but we must keep in mind, that for $m = n$ there is no
random choice in the communication regime.

The two mental models are very similar in the way that they both
use a threshold of \emph{bounded confidence} respectively
\emph{uncertainty}. We will call it $\eps$ according to the KH
model. The second parameter we use has its origins in the DW
model. We call it \emph{self-support} $\mu$. This parameter allows
formulating an averaging rule that is able to reproduce both
models. Since we specify a sub model of our general model the
following definition fulfills the properties required by
\ref{def:avru}.

\begin{definition}[averaging rule with $\eps$ and $\mu$]\label{recX1}
Let $C(t)\in\{0,1\}^n$ be a communication regime and $i\in\n$ be
an agent, $X(t)$ be an opinion profile and $0<\eps,\mu\in \left[0,1 \right]$.
Then we define the \emph{averaging rule} as
\begin{eqnarray*}
 \wei_{i}(j,X) &=&
 \left\{ \begin{array}{cl}
   \mu+\frac{1-\mu}{|I(i,X)|} \quad & \textrm{if } j=i  \\
   \frac{1-\mu}{|I(i,X)|} \quad & \textrm{if } j\in I(i,X) \textrm{ and } j\neq i  \\
   0 & \textrm{otherwise}
\end{array} \right. \\
&& \mbox{  with  } I(i,X) := \{j \in \n, \,|\, c_{ij} = 1 \textrm{
and } |X_{i} - X_{j}| \leq \eps \}
\end{eqnarray*}
(We omitted the specific time step for abbreviation.)
\end{definition}

Parameter $\mu$ is not the same as the parameter $\mu$ in the DW
model. To distinguish them we label the original parameter of the
DW model $\mu_D$.  If we set $\mu=1-2\mu_D$ and $m=2$, then our
model equals the DW model. The assumption of $0 < \mu_D \leq 0.5$
that is done in DW model corresponds to $0 \leq \mu <1$ in our
model. The parameter $\mu_D$ implements a tendency how much
another opinion influences the own opinion. Since our model
implements a multilateral communication this approach was not
easily extended. Should we assign a fixed cumulated influence to
all other influential opinions? We rejected this idea because we
aimed at a combination of the DW model and the KH model, and the
latter is incompatible with this idea. Instead we considered a
lower limit of the weight assigned to the own opinion. We call
this the self-support. On top of this value we add a value that
depends on the number of influential opinions. This allows a
smooth transition from the DW mental model to the KH mental model,
for which we have to set $\mu = 0$.

All together we have a unified basic model of continuous opinion
dynamics that allows for the analysis of the impact of different
numbers of communicating agents. It includes a special class of
communication regimes that is independent of social networks or in
other words a communication regime that has only the fully
connected network as its underlying social network. It
incorporates two important basic models that are applied in many
articles on continuous opinion dynamics with compromising agents,
i.e. the model of Deffuant and Weisbuch and the model of Krause
and Hegselmann. Actually both extremes happen in real social
structures. But also meetings with $m=3$ or 4 or 10 or 50 people
occur. Independent variables of our model are the number of agents
$n$, the number of communicating agents $m$, self-support $\mu$
and the bounded confidence parameter $\eps$. Actually the initial
profile $X(0)$ and the randomly chosen $m$-communication regime
$C(t)$ are also free variables of the model, but we will treat
them as endogenous random choices being equally distributed
within the defined bounds. Due to this random effect we are forced
to run many simulations with different randomly chosen $X(0)$ and
$C(t)$.

\section{Simulations} \label{sec:simulation}
For exploring the model presented in the last section we first define
three dependent variables: (1) final number of clusters, (2)
convergence time, and (3) last split time. Doing this, 
we show that we apply a simulation mechanism that is
equivalent to simulations of infinite many time steps. We then
report simulation results that explore the effect of changing the
number of communicating agents. Since our model also contains
the parameter self-support we additionally run simulations that
shed light on this parameter and -- more important -- how it
interacts with the number of communicating agents.

\subsection{Clusters, convergence and last split}

The most important outcome of an opinion dynamics process is the
final number of clusters. In many papers on opinion dynamics, 
a cluster usually refers to a set of agents that hold the same 
opinion. For continuous opinion dynamics this can be a difficulty
because the equality might only hold as a long-term limit. To
solve this problem we will define fully connected classes of agents
and show that for states with only fully connected classes, we can 
calculate the long-term limit of the dynamics.

It can be shown that each process in our general model covering the DW
and the KH model reaches a time step in which the agents split for all
further time steps into the same disjoint classes of agents for
which two conditions hold: (1) the distance of each agent in one class to each
agent in another class is greater than $\eps$ and (2) for one
class it holds that the maximal opinion in the class minus the
minimal opinion in the class is less or equal $\eps$. (For a proof
see \cite{Lorenz2003c}.) Thus for each class it holds that if some
agents of this class get in contact due to the communication
matrix then everybody trusts everybody directly. Thus we call this
class fully connected. Further, for a fully
connected class we can be sure that the opinions converge to the
same opinion, so to a cluster, but this can last in some cases
infinite time steps (due to $\mu$ and specific communication
regimes).

For continuous opinion dynamics reaching a state where all classes
are fully connected seems to be a convincing definition of
convergence. But the positions of the evolving clusters are not
unambiguously defined. Hence this definition of fully connected
classes makes an analysis of the distribution of clusters
difficult. To deal with this problem we apply the following lemma.

\begin{lemma} \label{lemma}
Let $X\in\R^n$ and let $A\in\R^{n\times n}$ be row-stochastic and
symmetric. Let $\bar{X}$ be the arithmetic mean of $X$, i.e.
$\bar{X} := \frac{1}{n} \sum_{i=1}^n X_i$. It holds $\bar{AX} =
\bar{X}$.
\end{lemma}
\begin{proof}
\begin{eqnarray} \nonumber
\bar{AX}=\frac{1}{n} \sum_{i=1}^n \sum_{j=1}^n a_{ij}X_j =
\frac{1}{n} \sum_{j=1}^n (X_j \underbrace{\sum_{i=1}^n a_{ji}}_{=
1})= \frac{1}{n} \sum_{j=1}^n X_j = \bar{X} &&\hspace{2.5cm}\Box
\end{eqnarray}
\end{proof}


Due to fully connectedness and definition \ref{recX1} we know for our model 
that for a system with only fully connected classes the confidence matrix $A$ is always row-stochastic and symmetric. The lemma states that the average of all
opinions of the fully connected class always stays the same, thus
even in the limit case. This allows to calculate the long-term limit of the dynamics.

All together we have a precise criterion when to stop a simulation 
instead of applying a heuristic. Our criterion stops at the point in time 
when we have only fully connected classes. The time needed to reach this 
state is the {\em convergence time}. We then calculate the long-term limit 
of the final {\em number of clusters}. Beside the number of clusters and convergence
time we also look at the {\em last split time}. This is the point
in time when the final set of classes emerges, but these
classes have not necessarily converged into fully connected
classes, i.e. only the first condition for fully connected classes holds. 
In simulations the last split time can only be determined
after determining the convergence time, because for not fully connected classes we
cannot easily determine whether it will split again or not. 
Despite last split time and
convergence time seem to be very similar, we will see in
simulations that they can show different behaviors.

\begin{figure}[htb]
  \includegraphics[width=\columnwidth]{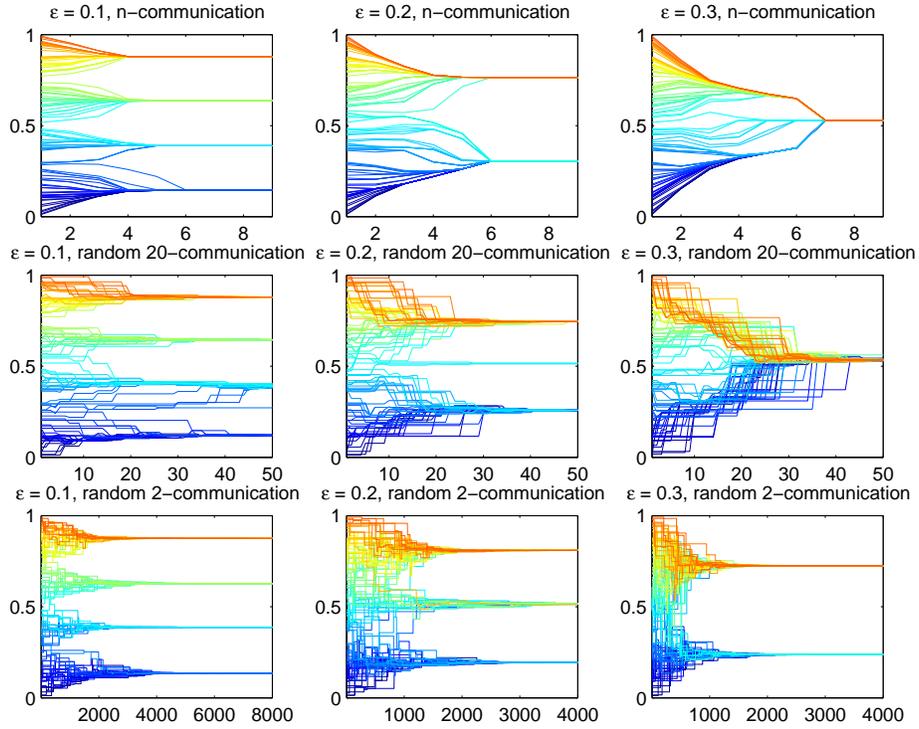}\\
  \caption{Examples: $100$ agents with randomly chosen initial opinions for $\eps$ from $\{0.1, 0.2, 0.3\}$ and for $m = n$ (KH model), $m=20$ and $m=2$ (DW model)}\label{fig}
\end{figure}

\subsection{Changing the number of communicating agents $m$} \label{subsec:changing}

For exploring the effect of changing the number of communicating
agents, $m$, we keep parameter $\mu$ constant at zero. For $m=n$ 
the model reduces to the KH
model and for $m=2$ we get the DW model with
$\mu_D=0.5$. The effect of changing parameter $\mu$ is explored
later in this paper.

To get a first idea, how $m$ may influence the dynamics we run nine
simulations. Figure \ref{fig} shows the nine processes of opinion dynamics, $\eps \in \{0.1,0.2,0.3\}$, $m \in \{2,20,100\}$), and all with the same initial profile of $100$ randomly chosen opinions. In each graphic, the
$x$-axis is time and the $y$-axis is the opinion of every agent. 
For low $\eps$ the agents form several clusters. For higher $\eps$
the number of clusters decreases. With even higher $\eps$ the
agents find consensus. All models show the same behavior but it
seems that the same $\eps$ causes slightly more clusters in the DW
model. Figure \ref{fig} shows and this also holds if the time needed to converge is transformed such that in every step the same number of agents adapt their opinions
(e.g. dividing $t$ by $\frac{100}{m}$ for the DW model) that for smaller values of $m$
it takes longer to converge. Similar graphs and the same
observations are reported in \cite{Hegselmann2004}.

To make the dynamics more visible we take a look at a very simple
example with three agents. Thereby we assume that for different
numbers of agents the fundamental micro behaviors will not change
significantly. Consider three agents with
opinions $0$, $0.5$, and $1$, all with $\eps = 0.5$ and any
self-support $\mu < 1$, e.g. $\mu = 0.4$. For the DW model
two clusters will emerge, either with opinions $0.0$ and $0.75$ or
with $0.25$ and $1.0$ (depending on the first communications). If
the extreme agents communicate they ignore each other. If the
middle agent communicates with an extreme agent it will adapt to
the extreme and leave the space where the other extreme agent could
influence it. For the KH model one cluster emerges with opinion
$0.5$. If we further increase epsilon above $0.5$ then the
probability to reach consensus increases for the DW model until
finally the probability reaches $1$. Between the extremes the
particular patterns of communication influence the probability of
convergence to one cluster.

All together we can state our first hypothesis: {\em The higher
the number of communicating agents, i.e. the bigger $m$, the less
the number of expected clusters and the less the convergence time
and last split time}.

\begin{figure}[htb] \centering \leavevmode
\includegraphics[width=\columnwidth]{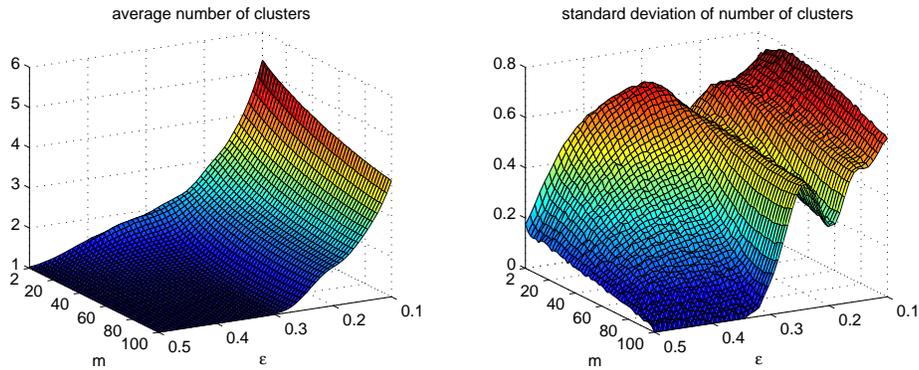} \\
\caption{Simulation results: epsilon and number of communicating
agents versus average number of clusters (left) and standard
deviation of the number of clusters (right)}
\label{fig01}
\end{figure}

\begin{figure}[htb] 
\centering \leavevmode
\includegraphics[width=12cm,height=4cm]{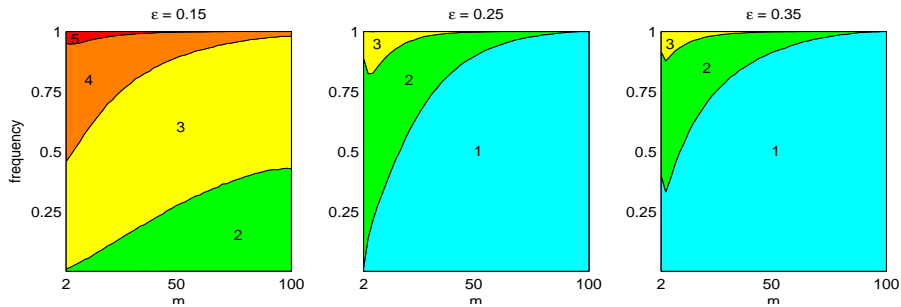} \\
\caption{Simulation results: Number of communicating agents versus 
probability of particular numbers of clusters for three instances of epsilon, i.e. $\eps = 0.15,0.25,0.35$ } 
\label{fig03}
\end{figure}

\begin{figure}[htb]
\centering \leavevmode
\includegraphics[width=\columnwidth]{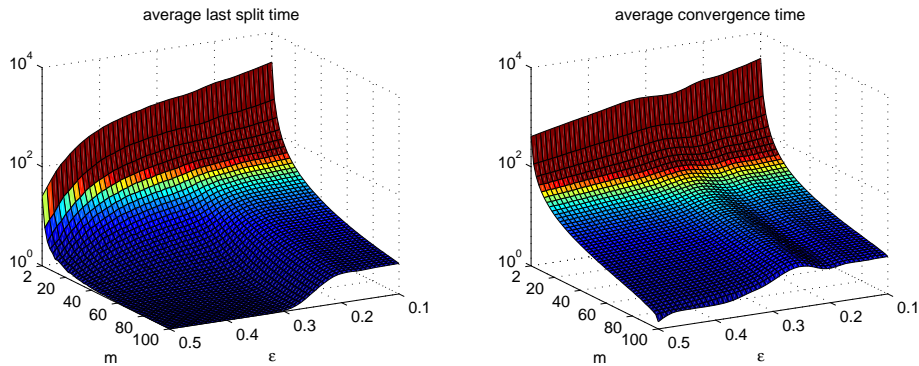} \\
\caption{Simulation results: epsilon and number of communicating
agents versus average last split time (left) and average
convergence time (right)} \label{fig02}
\end{figure}

To check our hypothesis we run simulations with $100$ agents
for different scenarios with $50$ stages for $m$ as well as $50$
stages for $\eps$. Every single scenario is simulated $10.000$
times with randomly selected initial opinion profiles. Figures
\ref{fig01}, \ref{fig02}, and \ref{fig03} visualize the results.

On average for
many different opinion profiles the number of clusters increases
the fewer agents communicate. In other words, for fewer
communicating agents we need more confidence, i.e. more
open-minded agents, to reach the same number of clusters. The
standard deviation of the number of clusters in many cases
decreases for bigger $m$; but for specific ranges of $\eps$ the
standard deviation decreases with an increase of $m$ for very
small $m$ (below 25) and decreases for bigger $m$; it has a maximum bigger than $2$ and less than $n$.

Figure \ref{fig03} illustrates these findings more detailed. For 
three different $\eps$, i.e. $\eps \in \{0.15, 0.25, 0.35\}$,
we plotted the probabilities of reaching a particular number
of clusters. The number of communicating agents has $30$ stages and
every setting was simulated $50.000$ times with randomly chosen
initial opinion profiles. For $\eps=0.35$ we observe that for 
an increasing $m$ the average number of clusters increases for 
very small $m$, while for more than $3$ agents it decreases as 
everywhere else. The same effect is not easy to see but can be 
found in Figure \ref{fig01} for epsilon between $0.25$ and $0.4$.
Despite this effect does not appear for $\eps =0.25$ and for the 
average number of clusters, a detailed inspection of \ref{fig03} 
shows that also in this case there is a specific behaviour fro small $m$.
The reason for this effect remains open for future research, but 
we can say that this effect gets smaller for bigger self-support 
$\mu$ (we omit the graphs here).

Figure \ref{fig02} shows the average last split time and the
average convergence time. It holds that the more agents
communicate the less steps agents need to converge.
The graphs show another very interesting behavior that holds for
the DW model as well as for the KH model: Despite there is a general tendency 
that the convergence time decreases with an in increasing $\eps$, there is a maximum
between $0.2<\eps$ and $\eps<0.3$. This does not hold for the last split time. 
Because this effect is not specific for 
changing $m$ we omit a further analysis, but we hypothize
that in the interesting area there is a tendency to polarize temporarily, but
convergence occurs in the long run. In such cases convergence time increases
significantly since convergence of two polarized groups take much time.

\subsection{Changing the self-support $\mu$}

For analyzing the effect of $m$ we set $\mu$ to zero,
because for this case both extremes, $m=2$ and $m=n$, represent
instances of the original DW model and KH model, respectively. If
we change $\mu$ then for $m=100$ we do not have the original KH
model anymore. We end up with a model that includes self-support
into the KH model. We now analyze how such changes affect the 
opinion dynamics. 

Since $\mu$ is based on the convergence parameter of the DW model
we can adapt an observation from \cite{DeffuantEtal2000} and hypothize
that {\em the smaller the self-support, the less is the convergence time};
 but, does it affect the average number of clusters?

Let us take a second look at the simple three-agent example
introduced in subsection \ref{subsec:changing}. Consider the
example for the DW model and $\eps$ bigger than $0.5$. The slower
the middle agent moves, i.e. the higher its self-support,
the lower is the probability that it "looses contact" to one of the 
extreme agents. And the lower this probability the higher is the 
probability that the
middle agent forces the extreme agents to reach consensus. Hence
we can hypothize that {\em there are cases where a higher
self-support decreases the expected number of clusters}. This
effect of changing $\mu$ could be strongest if extreme agents have to move a long
way while the middle agents have much time to move away; hence,
for big epsilon (where consensus is very likely) the effect could
be strongest. But on the other hand, the slower the extreme agents
move the higher is the probability that the middle agent looses
contact to an extreme agent, because the extreme agents do not get
closer to the middle agent quickly enough. The second effect could
also have its strongest impact for cases where the extreme agents
have to go a long way to reach a stable state, i.e. consensus. For
such conditions we might expect cases where an increase of the
self-support parameter increases the number of evolving clusters.
Hence we can also hypothize that {\em there are cases where the
number of clusters increases due to an increase in the
self-support $\mu$.} Both effect seem to have their strongest
impact for big epsilon. Simulations will show how these contrary
forces work.

\begin{figure}[b] 
	\centering \leavevmode
  \includegraphics[width=\columnwidth]{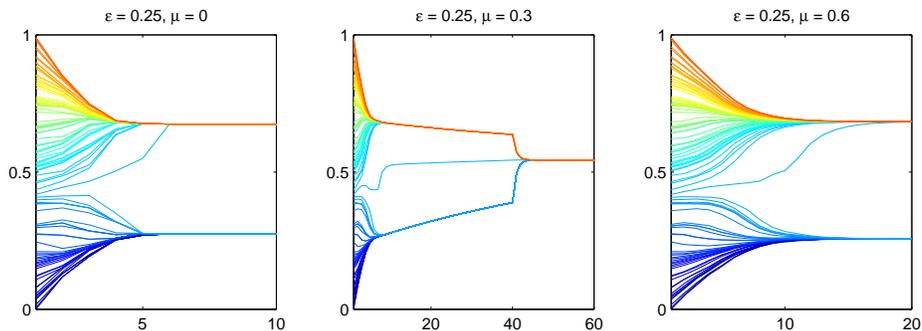}\\
  \caption{Examples for one initial opinion profile $X(0)\in\R^{100}$, $\eps = 0.25$ and $\mu = 0, 0.3, 0.6$}\label{muinKH}\label{fig:bspmu}
\end{figure}

To see that both effects are working see Figure \ref{muinKH}, where 
we consider an example with a fixed initial opinion profile of $100$
agents, $m=100$ (remember that there is no randomness in the $n$-random communication regime, i.e. the KH model), $\eps = 0.25$, and three stages for the 
parameter $\mu$. We see that raising $\mu$
from $0.0$ to $0.3$ takes the number of clusters from 2 to 1, but
raising $\mu$ further from $0.3$ to $0.6$ takes it back to 2 clusters.

\begin{figure}[t]
\centering \leavevmode
\includegraphics[width=\columnwidth]{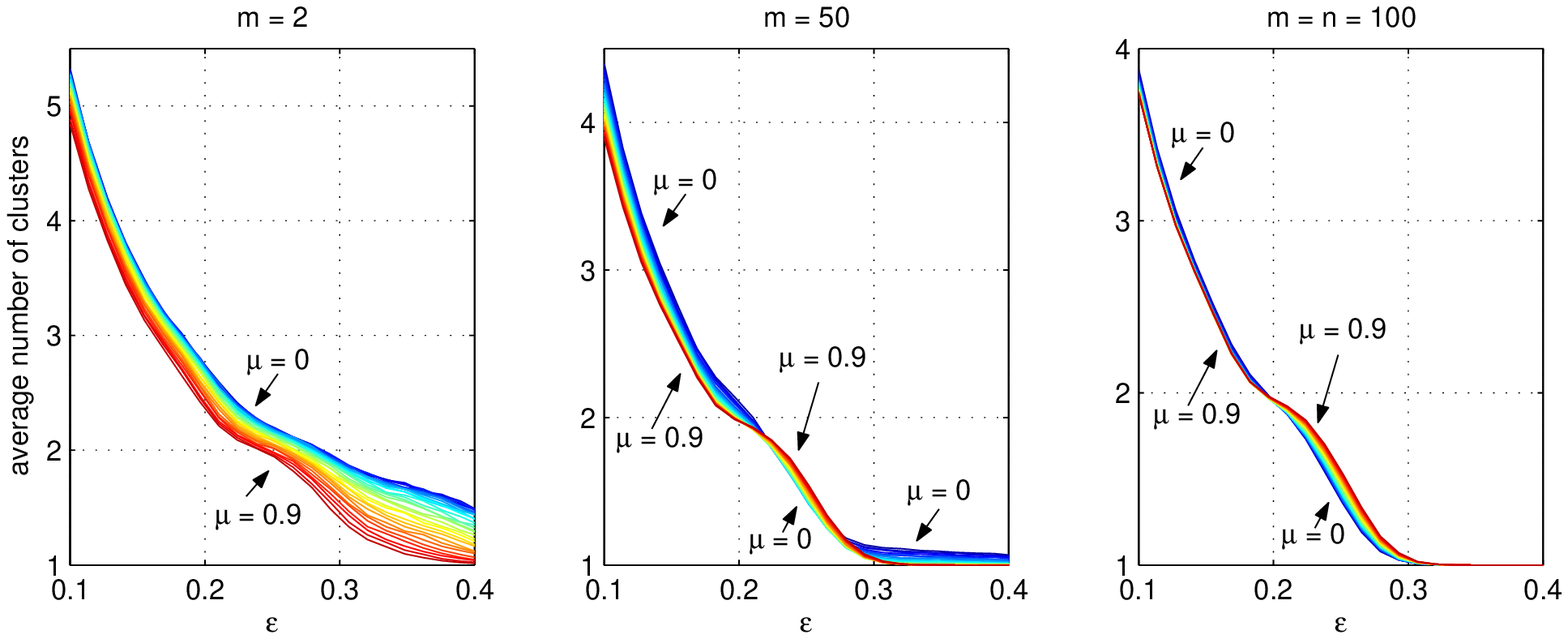} \\
\includegraphics[width=\columnwidth]{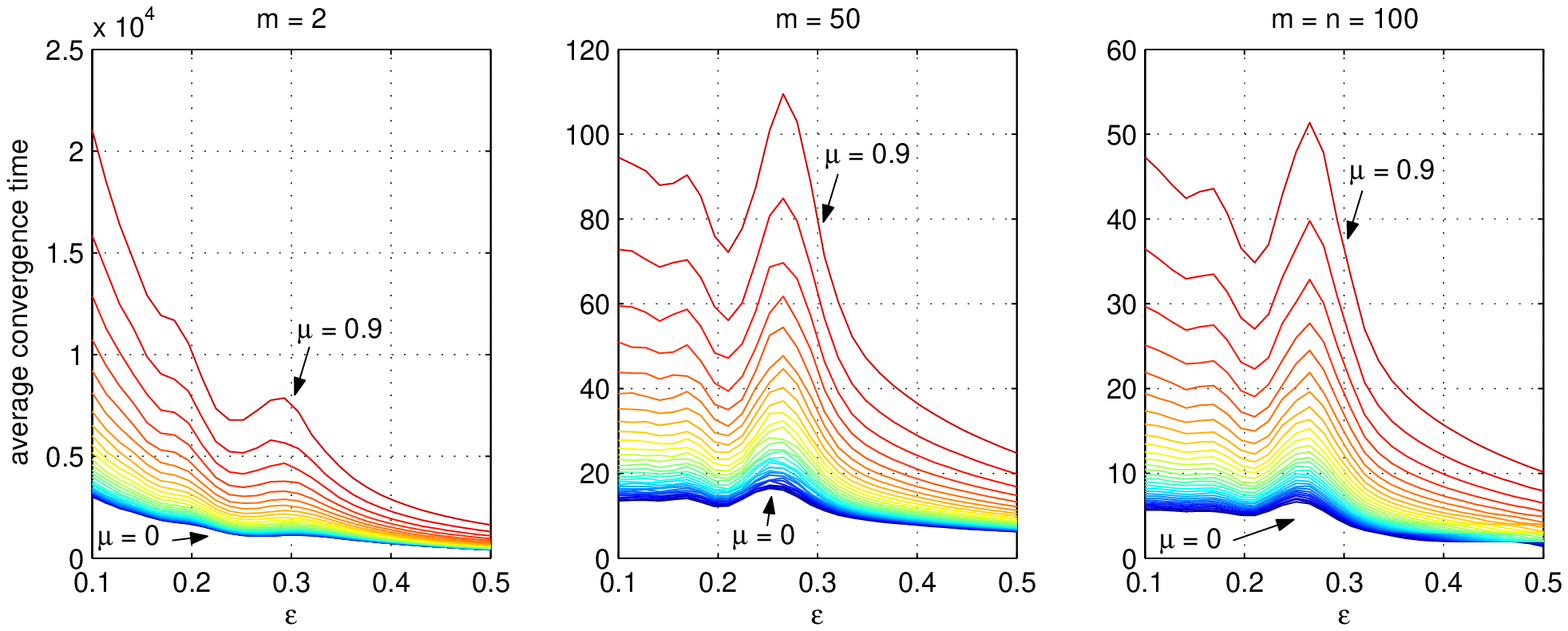} \\
\caption{Simulation results: epsilon versus average number of
clusters for different $\mu$ and for three different $m$. red
represents $\mu=0.9$ and blue represents $\mu=0.0$.} 
\label{fig07}
\end{figure}

Figure \ref{fig07} presents how $\mu$
affects on average the final number of clusters in the DW model (left) and in
the KH model (right), respectively. We run simulations for $100$ agents,
three stages of $m$, $30$ stages of $\eps$, and $30$ stages for $\mu$, 
while every setting was simulated $5.000$ times with random initial opinion profiles.

We can see that for the DW model ($m=2$) an increase of $\mu$ decreases
the number of clusters. For the KH model ($m=n=100$) we recognize a different
case. For small $\eps$ the direction of the effect is the same as
for the DW model (but the effect is much smaller); 
but for big $\eps$ it is inverse (but still the effect is smaller than for $m=2$). In fact, an
increase in $\mu$ for big $\eps$ increases the average number of
clusters.  

Regarding the convergence time
we can confirm our hypothesis and the observation reported in \cite{DeffuantEtal2000}, i.e. the bigger $\mu$, i.e. the smaller $\mu_D$, the longer the agents need to converge.

However, our results regarding the number of clusters seem to be in partial contradiction 
with one observation mentioned in \cite{DeffuantEtal2000}: "$\mu$ and $N$ [=$n$
in our model] only influence convergence time and the width of the
distribution of final opinions (when a large number of different
random samples are made)." The difference might be due to the fact
that we do not exclude "wings", i.e. asymmetric
peaks with a vertical bound of either 0 or 1, from our data set.
Independent of this, we see that the effect of $\mu$ changes when
the communication regimes changes. Hence, it might be interesting
how the opinion dynamics change if networks are introduced as part of
communication regimes.

Until now we have only analyzed averages of many different initial
opinion profiles and discussed the major effects. 
In this paragraph we briefly summarize observations
that are not dominant or that appear only for specific or fixed initial opinion profiles. Due to page 
limitations we do not explain these effects in detail. For changing the number of communicating agents $m$ for given epsilon and given self-support we observed two minor effects: For some epsilon between $0.28$ und $0.4$ a change from $m=2$ to $m=3$ or $m=4$ increases the number of clusters, while usually it decreases. 
Also, for some initial opinion profiles the change from $m=n-1$ to $m=n$ causes some non-monotony, because
by this change we eliminate randomness in the communication regime. For changes in epsilon for $m=n$ and $\mu=0$ (KH model) non-monotonic behavior is already reported \cite{Lorenz2003c}; however, from our simulations we can report this also for the DW model for specific initial opinion profiles (e.g. equidistant initial opinions). Figure \ref{fig:bspmu} shows that non-monotonic behavior also appears for changes in $\mu$ for specific initial opinion profiles. All together, we can summarize that the average behavior of our model seems to be relatively smooth and clean, but for specific parameter settings and specific initial opinion profiles there are qualitative deviations from the average behavior.
 
\section{Conclusion and Outlook}

This article contributed to the literature on continuous opinion
dynamics in four ways. First, we provided a general model of
continuous opinion dynamics that distinguishes the mental model
from communication regimes. We discussed how ideas like social
networks and existing models are related to the idea of
communication regimes. Second, based on the general model we
developed a more specific model that contains two of the most
influential models of continuous opinion dynamics. 
The crucial parameter that distinguishes these two models
is the number of agents that communicate in one step, i.e. the
communication regime. The unifying model allows an exploration of
the space between the extremes. We were able to show that an
increase in the number of communicating agents mostly leads to a 
decrease of the average number of emerging clusters. There is a small
difference in the mental models of the Deffuant and Weisbuch model and
the Krause and Hegselmann model. The first introduces a parameter that
we picked up and that we have implemented similar as
self-support. We were able to show that there is an interaction
effect between the number of communicating agents the
self-support. A fourth contribution is not related to model
analysis but to the implementation of simulations. We pointed out
that for the unifying model, which covers the DW model as
well as the KH model, there is a theoretically
driven condition that allows to stop a simulation and to calculate
the simulation result for a simulation with infinite many steps.
Having this, we are able to "simulate" infinite many steps.

Several question remain open for future work. 
We need to explore the effect of the number $n$ of
agents in the system and the interaction effects between all
parameters $m$, $n$, $\mu$, and $\eps$. We also have to explore
heterogeneous societies. For implementations we want to extend our
stop criteria to implementations of opinion dynamics on social
networks.

\bibliographystyle{plain}

\end{document}